# Introducing students to the culture of physics: Explicating elements of the hidden curriculum

Edward F. Redish[1]

[1]Depts. of Physics and Curriculum & Instruction, University of Maryland, College Park, MD 20742, USA

**Abstract.** When we teach physics to prospective scientists and engineers we are teaching more than the "facts" of physics – more, even, than the methods and concepts of physics. We are introducing them to a complex culture - a mode of thinking and the cultural code of behavior of a community of practicing scientists. This culture has components that are often part of our hidden curriculum: epistemology - how we decide that we know something; ontology - how we parse the observable world into categories, objects, and concepts; and discourse - how we hold a conversation in order to generate new knowledge and understanding. Underlying all of this is intuition – a culturally created sense of meaning. To explicitly identify teach our hidden curriculum we must pay attention to students' intuition and perception of physics, not just to their reasoning.

## INTRODUCTION

Learning a complex field of knowledge such as physics is a more complex process than we often give it credit for – because it is so natural to us. Even though we may have spent more than twenty years in school and many more learning to be experts, we have a strong tendency <u>not</u> to think explicitly about learning. Rather, we have "phenomenological primitives (p-prims)" of learning and folk psychology. But to understand our students and their difficulties in mastering physics, we have to get beyond our p-prims about their thinking and do the science of understanding their thinking processes.

Typically, when we prepare our syllabi, we select the content we intend to "cover". But what we really want our students to learn is more than just a set of facts: it's a way of thinking – the manner and "adaptive expertise" of the professional. The "hidden curriculum" is made up of those elements that are a part of learning our subject that we expect our students to learn without our being explicit about them – the items that are supposed to "come along for the ride".

In order to go deeper than just "that-looks-good-to-me!" heuristics, we have to do some deep thinking about the nature of professionalism in physics. We need to develop insight into how the culture of physics affects our cognitive processing and our thinking about the physical world. Some issues have been explicitly studied by the PER community, for example:

- Epistemology – What is accepted as evidence for believing a particular result? [1]
- Ontology – What kind of "things" do we talk about and what is their nature? [2]

But there are others that I think we need to be more explicit about. One that we often ignore is *intuition*. In order to consider this element, let's think deeply about both how people think in general and how professionals think. We'll look at the hidden curriculum through two lenses: the cognitive and the socio-cultural.

## THE HIDDEN CURRICULUM THROUGH A COGNITIVE LENS

Since we tend to learn without thinking about it explicitly, we need to deconstruct expert thinking. [3] This helps us make sense both of why the hidden curriculum is hidden and what elements we need to begin paying attention to.

### Why the Hidden Curriculum is Hidden

Learning involves a number of cognitive developments: building links and associations, compiling associated elements into chunks that can be used as single elements in working memory (and opened up when needed), and developing framings – learning when particular bits of knowledge and activities are relevant. These items are discussed in a number of papers (see [4], section 4 and refs. therein). They help explain why things that seem trivial to an expert may be difficult for a student and why the process of developing expertise remains "hidden". A critical element in under-standing both of these issues is intuition. Its cognitive content is suggested by two parables.

## Two Parables of Achilles

The first parable comes from a book on the subject of cognitive linguistics. [5] In this book, Gilles Fauconnier and Mark Turner consider how we make meaning of language and the detailed ways in which we construct new ways of thinking out of old. They are not trying to make sense of the formal elements of language, but rather of what goes on invisibly "under the hood". Here is a selection from their preamble.

> *We live in the age of the triumph of form. In mathematics, physics, music, the arts, and the social sciences, human knowledge and its progress seem to have been reduced in startling and powerful ways to a matter of essential formal structures and their transformations…. The axiomatic method rules, not only in mathematics but also in economics, linguistics, sometimes even music.*
>
> *On the other hand, common sense tells us that form is not substance: The blueprint is not the house, the recipe is not the dish, the computer simulation of weather does not rain on us. When Patroclos donned the armor of Achilles to battle the Trojans, what the Trojans first saw was the spectacular armor, and they naturally assumed it was Achilles, and were terrified, and so the armor by itself looked as if it was turning the battle. But it didn't take long for the Trojans to discover that it was just Achilles's armor, not Achilles himself, and then they had no pity…. Clearly the miracles accomplished by the armor depend on the invisible warrior inside.* [5]

Fauconnier and Turner go on to flesh out the Achilles metaphor. Our formal linear reasoning is like the armor – powerful and enhancing; but without the "inner Achilles" – our parallel processing "recognition software" – the formal reasoning is useless. Some psychologists work very hard to try to build computational models of recognition, but do not yet seem to be able to help us at the phenomenological level. For now, I suggest we need to think of "intuition" as something distinct from formal reasoning but enabling it.

This point is strengthened further by a second parable of Achilles, Lewis Carroll's "What the Tortoise Said to Achilles". In it the tortoise undermines Achilles' application of formal logic. Here's an excerpt. [6]

> *Tortoise:* "Well, now, let's take a little bit of the argument in that First Proposition -- just two steps, and the conclusion drawn from them. Kindly enter them in your notebook. And in order to refer to them conveniently, let's call them A, B, and Z: --
>
>    *(A) Things that are equal to the same are equal to each other.*
>    *(B) The two sides of this Triangle are things that are equal to the same.*
>    *(Z) The two sides of this Triangle are equal to each other.*
>
> *T:* Readers of Euclid will grant, I suppose, that Z follows logically from A and B, so that any one who accepts A and B as true, must accept Z as true?"
>
> *Achilles:* "Undoubtedly! The youngest child in a High School -- as soon as High Schools are invented, which will not be till some two thousand years later -- will grant that."
>
> *T:* "And might there not also be some reader who would say 'I accept A and B as true, but I don't accept the Hypothetical ' [i.e., proposition Z]?"
>
> *A:* "Certainly there might. He, also, had better take to football."
>
> *T:* "And neither of these readers," the Tortoise continued, "is as yet under any logical necessity to accept Z as true?"
>
> *A:* "Quite so," Achilles assented.
>
> *T:* "Well, now, I want you to consider me as a reader of the second kind, and to force me, logically, to accept Z as true."
>
> *A:* I'm to force you to accept Z, am I?" Achilles said musingly. "And your present position is that you accept A and B, but you don't accept the Hypothetical --"
>
> *T:* "Let's call it C," said the Tortoise."
>
> *A:* "-- but you don't accept
>    *(C) If A and B are true, Z must be true.*"
>
> *T:* "That is my present position," said the Tortoise.
>
> *A:* "Then I must ask you to accept C." [6]

You can see where this is going. It is making the point that making the connection between properties A, B, and Z does not close the argument but logically must be considered as a proposition itself. This is rather startling – especially to a theoretical physicist such as myself. It's Zeno's paradox applied to logic. The discussion clarifies the role of our "inner Achilles". Even in formal proof our recognition software steps in to cut off the infinite chain. In other words, "Oh, I get it!" is NOT an element of formal logic.

This explains something that I've often seen in my classes. I will be doing a derivation on the board when some student asks, "How did you get that step?" Looking at it, it appears totally obvious to me. Looking at it again, I realize that I have carried out a number of steps at once – multiplying by 2, moving something to the other side of the equation, etc., and that *I* could recognize that the two results were the same without thinking about it. To get my students to recognize their equivalence I had to explicate a number of steps. It becomes clear that *these intuitive recognitions are cultural norms, learned as we become physicists.* They play a role at different level than the formal, but they play a critical role even in interpreting the formal.

Applying appropriate intuitions is not automatic, even for professionals. Consider the following.

### An Example from Problem Solving

An interesting example is provided by a problem and solution from a recent (highly popular) calculus-based physics text. Since the problem I am about to cite contains an error, I will not tell you what book it comes from. Every text contains errors, and I don't mean to beat up on a particular author for missing something in the thousands of problems contained in a modern text. The problem is shown below.

> On a hot 35 C day, you perspire 1.0 kg of water during your workout.
> (a) What volume is occupied by the evaporated water?
> (b) By what factor is this larger than the volume occupied by the liquid water?

*Fig. 1: A problem from a calculus-based introductory physics text.*

This seems quite strange. We know a lot about gases, in particular, that they expand to fill space. Is the answer to (a), "it depends on the size of the room you are working out in"? To see what they had in mind we look at the solution manual. To help students, the text suggests a rubric for approaching problem solving.

> *Model*! –Make simplifying assumptions.
> *Visualize*! – Draw a pictorial representation.
> *Solve*! – Do the math.
> *Assess*! – Check your result has the correct units, is reasonable, and answers the question.

This is good advice, isn't it? Here's the solution in the manual. Note that it claims to follow the rubric.

> *Model*: –Assume the evaporated water is an ideal gas with a molar mass of 18 g/mole. Assume the pressure is 1 atm = 101.3 kPa.
> *Visualize*: We are given
> $$T = 35\ C + 273 = 308\ K.$$
> $$n = 1000\ g(1\ mol/18\ g) = 55.6\ mol.$$
> *Solve*:
> (a) $$pV = nRT \Rightarrow V = \frac{nRT}{p}$$
> $$= \frac{(55.6\ mol)(8.31\ J/mol \cdot K)(308\ K)}{101.3\ kPa} = 1.4\ m^3$$
> (b) In the liquid state $\rho = 1000\ kg/m^3$
> (a simple calculation yields a factor of 1400).
> *Assess*: Gases really do take up a lot more volume than the equivalent mass of a liquid!

Ouch! The difficulty is that the answer in the solution manual fails to treat the problem intuitively – to *tell the story of the problem.* The steps of the rubric are gone through formally – but without the activation of a sense-making intuition about molecules – and even about the nature of gases.

Tying the analysis to a rubric – a formal set of mapped rules (the armor) does not help if it does not also activate an intuitive sense of meaning by tying the problem to all we know and recognize about a system. Again, my point is not to chastise the authors of this particular problem but to point out that even professionals can produce nonsense if they mistake the armor for the inner Achilles.

Intuition has many components: the identification of identity – to determine when things are supposed to represent the same thing (as in being able to follow formal proof); and the making of meaning – placing a problem in a broader context by linking to the many things we know about our subject and about the world.

## THE HIDDEN CURRICULUM THROUGH A SOCIO-CULTURAL LENS

What I am calling intuition relies on our entire experience of the world, but it also involves strengthening our recognition software through professional knowledge and experience. The intuition of a physicist does not belong to an individual alone, but arises as an emergent phenomenon from his or her interactions with others in the community of physicists and from the educational experiences we create for our students. This has powerful implications when we consider interdisciplinary issues such as getting training for our majors from service departments such as math – or when we ourselves provide training for majors in other disciplines such as engineering or biology.

### Math in Physics

If many of our critical hidden curricular elements are community driven, when we go to different communities for service courses we can get something different from what we want.

Here's a problem from a calculus final exam at my university.

> The population density of trout in a stream is
> $$r(x) = 20\frac{1+x}{x^2+1}$$
> where $r$ is measured in trout per mile and x is measured in miles. $x$ runs from 0 to 10.
> (a) Write an expression for the total number of trout in the stream. Do not compute it.
> (b)…

*Fig. 2: A problem from a final exam in an introductory calculus class.*

I suspect that most physicists will be troubled, as I am, by the lack of concern for units. Not only is the "20" a different kind of object than the "1", but the "1" on top of the fraction and the "1" on the bottom are different kinds of objects! (One is a length, the other an area.) If one of my students wrote an expression like that on my exam, they would get a 0 for writing nonsense!

A second problem is that of parameters. On the final calculus exam I looked at, every problem had two symbols – an independent variable and a dependent one. All of the other glyphs on the exam were numbers or math symbols; there was not a single parameter. This avoidance of parameters is endemic in calculus. In a typical calculus text containing thousands of equations, you will find almost none containing parameters. Yet in physics, understanding parametric dependence – considering limiting cases, for example – is a critical skill. In a typical physics text, you will find few equations that do *not* contain numerous parameters. Our first equations in kinematics will contain half-a-dozen different letters – and confuse students terribly.

My third example gets at the heart of the hidden ways that we use math differently in physics from the way mathematicians do. Do we ever need our students to be able to create an epsilon-delta proof? In fact, I claim that in physics, the "derivative" is an approximation to a physical quantity that is *not* correctly described by a limiting process. We consider a velocity, for example, as a ratio of small changes. If we try to make $\Delta t$ too small, our $\Delta x$ will start losing its smoothness. We will see the bumps arising from the fibers on a wooden track, the vibrations due to Brownian motion, and eventually, the loss of definition of position at all due to quantum mechanics. I tell my students that when I write *dx/dt* the "*d*"s simply mean a $\Delta$ that is smaller than any scale I want to consider.

These distinctions illustrate hidden differences between the epistemological assumptions the cultures of physics and math make in the use of equations. If we fail to be aware of these differences – and to communicate them clearly to our mathematical colleagues, they can cause our students considerable difficulties.

### Physics in Biology

My math example shows how the mathematicians' hidden epistemological assumptions about their nature of knowledge sometimes have conflicts with ours in ways that cause our students trouble. Similarly, when we provide service courses, we need to understand how students in other disciplines use what we teach.

Many groups are now exploring how to better teach physics to biology students. Our traditional approach cuts down the course for engineers. But do biologists ever need to calculate projectile motions? Why should we teach thermo using heat engines? No biological organism makes its metabolism on temperature differences. Why do we essentially *never* mention chemical energy?

Beyond the content, what most biologists *do* with physics is often very different from what physicists or engineers do with it. The "hidden curriculum epistemology" of physics as used in biology embeds the physics into highly complex webs of knowledge about biological systems. In physics, we tend to always go to the simplest example and to build intuitions about them, using those intuitions as the bones on which to put the flesh of more complex examples. This doesn't work in biology where the physics acts more as constraints (energy conservation, second law of thermodynamics, charge conservation, …) and by describing functional parametric dependence that can be used by evolution in different way by different organisms.

### CONCLUSION

We often look at student failures and try to fix them by creating linear algorithms – as if the students were computers to be programmed. This sometimes works – because the inner Achilles comes along for the ride. I conjecture that many of the excellent PER reforms (Tutorials, Group Problem Solving, Workshop Physics [7]) work because they give students free rein to develop their inner Achilles.

But if want to learn how to help our students in a variety of communities build an adaptive and flexible expertise, we are going to have to pay more explicit research attention to understanding and modeling intuition development – from the intersection of a cognitive and socio-cultural perspective.

### ACKNOWLEDGMENTS

This work was supported by NSF grants 05-24987 and 09-19816.